\documentclass[man,floatsintext, a4paper]{apa6}

\usepackage[american]{babel}
\usepackage[utf8]{inputenc}
\usepackage{graphicx}
\usepackage{csquotes}
\usepackage{amsmath,mathtools}
\usepackage{amsthm}
\usepackage{amssymb}
\usepackage{bm,array}
\usepackage{latexsym}
\usepackage{color}
\usepackage{blkarray}
\usepackage{multirow}

\usepackage[]{hyperref}
\usepackage[style=authoryear,sorting=nyt,uniquename=false,sortcites=true,natbib=true,backend=bibtex]{biblatex}

\AtEveryBibitem{%
   \clearfield{labelmonth}
   \clearfield{month}
   \clearfield{note}
   \clearfield{number}
   \clearfield{url}
   \clearfield{urlday}
   \clearfield{urlmonth}
   \clearfield{urlyear}
}

\providecommand{\e}[1]{\ensuremath{\times 10^{#1}}}

\usepackage{tikz}
\usepackage{tikz-qtree}
\usetikzlibrary{trees}

\newcommand\numberthis{\addtocounter{equation}{1}\tag{\theequation}}

\theoremstyle{definition}

\title{Adjusted Priors for Bayes Factors Involving Reparameterized Order Constraints}

\shorttitle{Reparameterization of order constraints}

\twoauthors{Daniel W. Heck}{Eric-Jan Wagenmakers}
\twoaffiliations{University of Mannheim}{University of Amsterdam}

\leftheader{Heck \& Wagenmakers}

\authornote{
Daniel W. Heck, School of Social Sciences, Department of Psychology, University of Mannheim, Germany. Eric-Jan Wagenmakers, University of Amsterdam, Department of Psychological Methods, Amsterdam, the Netherlands.

This work was supported in part by the University of Mannheim's Graduate School of Economic and Social Sciences funded by the German Research Foundation. Code and data associated with this paper have been lodged with the Open Science Framework at \url{https://osf.io/cz827/}. 

Correspondence concerning this article should be addressed to Daniel W. Heck, Department of Psychology, School of Social Sciences, University of Mannheim, Schloss EO 254, D-68131 Mannheim,  Germany. E-mail: dheck@mail.uni-mannheim.de; Phone: (+49) 621 181 2145.

}

\abstract{
Many psychological theories that are instantiated as statistical models imply order constraints on the model parameters. To fit and test such restrictions, order constraints of the form $\theta_i  \leq  \theta_j$ can be reparameterized with auxiliary parameters $\eta\in [0,1]$ to replace the original parameters by $\theta_i = \eta\cdot\theta_j$. This approach is especially common in multinomial processing tree (MPT) modeling because the reparameterized, less complex model also belongs to the MPT class. Here, we discuss the importance of adjusting the prior distributions for the auxiliary parameters of a reparameterized model. This adjustment is important for computing the Bayes factor, a model selection criterion that measures the evidence in favor of an order constraint by trading off model fit and complexity. We show that uniform priors for the auxiliary parameters result in a Bayes factor that differs from the one that is obtained  using a multivariate uniform prior on the order-constrained original parameters. As a remedy, we derive the adjusted priors for the auxiliary parameters of the reparameterized model. The practical relevance of the problem is underscored with a concrete example using the multi-trial pair-clustering model.
}

\keywords{Inequality constraints, model selection, MPT modeling, model complexity, encompassing prior.}

\begin{document}
\maketitle



Linear order constraints of the form $\theta_i\leq\theta_j$ are important for many substantive hypotheses in mathematical psychology \citep{iverson2006essay,hoijtink2011informative}. In the present paper, we are mainly concerned with order constraints in multinomial processing tree (MPT) models \citep{erdfelder2009multinomial, batchelder1999theoretical}, a class of cognitive models that accounts for response frequencies by a finite number of underlying processes. In MPT models, the parameters $\bm \theta$ represent conditional and unconditional probabilities that specific cognitive states are entered. Hence, order constraints are necessary to test whether some of the underlying processes occur more often under specific conditions \citep{knapp2004representing}. 

Linear order constraints on MPT parameters can easily be represented using auxiliary parameters, resulting in a new, reparameterized MPT model. Importantly, such an order-constrained model makes more specific predictions than the original, unconstrained model and is less complex even though both models have the same number of free parameters \citep{myung2000importance,vanpaemel2009measuring}. Such a difference in complexity is taken into account by the  Bayes factor, the standard Bayesian model selection tool that is able to quantify the evidence in favor of the order constraint \citep{klugkist2005inequality, hoijtink2011informative}.

In the present paper, we show that the combination of both methods --Bayes factors for order-constraints in a reparameterized MPT model-- is prone to an inadvertent choice of prior distributions that do not match the actual prior beliefs.  Specifically, we illustrate the substantial difference between results obtained under a uniform prior on the original, constrained parameter space and those obtained under a uniform prior on the auxiliary parameters. Whereas the former meets the intuition that all admissible, original parameter vectors are equally likely, the latter leads to an overweighting of small values of the original parameters. We derive adjusted priors for the auxiliary parameters that imply a uniform prior on the original parameters and underscore the importance of this adjustment in case of the pair-clustering model \citep{batchelder1980separation}. Despite our focus on MPT models, the present paper highlights the importance of carefully choosing priors for reparameterized models in general.

\subsection{Reparameterization of Order Constraints in MPT Models}
 
When maximizing the likelihood of a model, many optimization procedures require that each parameter varies independently on a fixed range (e.g., between zero and one). However, if a model contains order constraints of the form $\theta_i \leq \theta_j$, the admissible values for one parameter depend on the values of the other parameters. As a practical solution, models are often reparameterized into statistically equivalent models with independent parameters. For instance, within a bounded two-dimensional parameter space $\bm \theta \in [a,b] \times [a,b]$, the constraint $\theta_1 \leq \theta_2$ can be implemented by auxiliary parameters defined as $\eta_2=\theta_2$ and $\eta_1 = \theta_1 / \theta_2$ \citep{knapp2004representing}. The equivalent expression $\theta_1=\eta_1\theta_2$ provides an intuitive interpretation of $\eta_1$ as a `decrease parameter' between zero and one. The reparameterized model with two independent parameters $(\eta_1, \eta_2) \in [0,1] \times [a,b]$ is statistically equivalent to the original model because both models imply the same set of probability distributions over the data. 

MPT models are special with regard to linear order constraints, because the reparameterization results in a new, less flexible MPT model \citep{knapp2004representing,klauer2015parametric}. In other words, the use of auxiliary parameters $\eta \in [0,1]$ to implement order constraints is equivalent to adding one or more branches in the graphical representation of a processing tree. This result facilitates testing order constraints with standard software for MPT models \citep[e.g.,][]{moshagen2010multitree, singmann2013mptinr} or hierarchical extensions \citep[e.g.,][]{klauer2010hierarchical,smith2010betampt}. Given these benefits, it is not surprising to find many applications of this approach, for instance, concerning pair-clustering in memory \citep{riefer2002cognitive,broder2008different}, outcome-based strategy selection in decision making \citep{hilbig2014generalized}, recognition memory \citep[][]{kellen2015signal}, and stochastic dominance of binned response time distributions \citep{ heck2016extending}.

\subsection{The Bayes Factor}

Bayesian model selection is based on the Bayes factor \citep{kass1995bayes,jeffreys1961theory}, defined as the odds of the data under models $\mathcal M_0$ vs. $\mathcal M_1$:
\begin{equation}
\text{B}_{01} = \frac{p(\bm y \mid \mathcal M_0)}{p(\bm y \mid \mathcal M_1)}.
\end{equation}
In order to obtain the marginal likelihood of the data $\bm y$ given the model $\mathcal M_0$, the likelihood $f_0$ is integrated over the parameter space $\Theta$ weighted by the prior $\pi_0$,
\begin{equation}
\label{e.marginal}
p(\bm y \mid \mathcal M_0) = \int_{\Theta} f_0(\bm y \mid \bm \theta) \pi_0(\bm \theta ) \, \mathrm d \bm \theta.
\end{equation}
Mathematically, this is the prior-weighted average of the likelihood \citep{wagenmakers2015power}, which explains the substantial impact of the prior on model selection. The Bayes factor can directly be interpreted as the evidence in favor of one model compared to another and is ideally suited to evaluate order constraints, which restrict the volume of the parameter space $\Theta$ \citep{klugkist2005inequality,klugkist2007bayes}. If an order constraint restricts the parameters to be in the admissible subspace $\Theta' \subset \Theta$, the integral in Eq.~\ref{e.marginal} will be large if the weighted likelihood on the subspace $\Theta'$ is high, but small if the weighted likelihood on the subspace $\Theta'$ is low. 

When testing order constraints on bounded parameters, an important and common prior is the uniform distribution, which has a constant density on the restricted parameter space $\Theta'$ and puts equal probability mass on all admissible parameter vectors \citep{klugkist2005inequality}. Note that this prior is theoretically appropriate to model, for instance, equally-likely error probabilities \citep{lee2016bayesian,myung2005bayesian}, but might be inappropriate for other applications (e.g., guessing probabilities expected to be around 50\%). Importantly, this multivariate uniform distribution refers to the original parameters $\bm \theta$. If order constraints are implemented by reparameterization, uniform distributions on the auxiliary parameters $\bm\eta$ will in general imply a nonuniform distribution on the original parameters  $\bm\theta$ \citep{fisher1930inverse,jaynes2003probability}. Hence, it is necessary to use adjusted priors for the auxiliary parameters, which we derive in closed form for linear order constraints of the type $\theta_1 \leq \theta_2 \leq  \dots\leq \theta_P$ below. Note that our results are not limited to MPT models but do also apply to other models with bounded parameter spaces (up to scaling constants).


\section{Reparameterized Order-Constraints in a Bayesian Framework}

To illustrate the importance of priors for reparameterized, order-constrained models, we use the product-binomial model, a trivial MPT model:
\begin{equation}
f^\theta(\bm y \mid \bm \theta)=\prod_{i=1}^{P} \text{Bin}(y_i, n_i, \theta_i).
\label{e.productbin}
\end{equation}
This model assumes that the frequencies $y_i$ across $P$ conditions are independent binomial random variables with success probabilities $\theta_i$ and number of draws $n_i$. In this simple model, we test the hypothesis that the $P$ conditions are ordered with increasing success probabilities, which is equivalent to the order constraint $\theta_1 \leq \theta_2 \leq  \dots\leq \theta_P$. Based on the approach by \citet{knapp2004representing} outlined above, we use the auxiliary parameters $\bm \eta$ defined recursively by $\eta_P=\theta_P$ and $\eta_i = \theta_i / \theta_{i+1}$ representing the relative decrease from $\theta_{i+1}$ to $\theta_i$ for all $i=1,\dots,P-1$. By replacing the original parameters, the likelihood of the reparameterized model becomes
\begin{equation}
f^\eta(\bm y \mid \bm \eta)=\prod_{i=1}^{P} \text{Bin}(y_i, n_i, \prod_{j=i}^P \eta_j).
\end{equation}
Note that this is just one of many possible reparameterizations for MPT models \citep[i.e., Method A of][Appendix~\ref{a.methodb} extends our results to Method B]{knapp2004representing}. 

The original, order-constrained model\footnote{We use superscripts for the parameterization and subscripts for the prior. Hence, the models $\mathcal M^\eta_u$ and $\mathcal M^\eta_a$ share the parameter $\bm\eta$ but assume a uniform and an adjusted prior (derived below), respectively.} $\mathcal M^\theta_u$ has a uniform prior on the substantive parameters $\bm\theta$ and thus a constant density on the admissible parameter space,
\begin{equation}
\pi_u^\theta (\bm \theta) = P! \cdot \mathcal I_{\{\theta_1\leq \dots \leq \theta_P\}}(\bm \theta),
\end{equation}
where $\mathcal I_{\{\theta_1\leq \dots \leq \theta_P\}}$ is the indicator function, which is one if $\theta_1\leq \dots \leq \theta_P$ and zero otherwise. This model is henceforth termed `balanced model' because the original, substantive parameter values are assigned equal prior mass.

In contrast, the  reparameterized model $\mathcal M^\eta_u$  assumes independent uniform priors on the auxiliary parameters $\bm \eta$ and thus a constant density,
\begin{equation}
\pi_u^\eta(\bm \eta) =\prod_{i=1}^{P} \mathcal I_{[0,1]}(\eta_i).
\end{equation}
We call this the `unbalanced model,' because --as shown in the next section-- the original parameters are not equally weighted. Note that this reparameterized model can also be implemented without the explicit use of auxiliary parameters $\bm \eta$ by using conditional uniform priors on the original parameters, that is, assuming  that the constrained parameters $\theta_i$ have a constant density on the range $[0,\theta_{i+1}]$.\footnote{In the software JAGS \citep{plummer2003jags}, this is done by \texttt{theta[i]  $\mathtt\sim$ dunif(0,theta[i+1])}.} 

\subsection{Implied Prior Distributions on the Original Parameters}
\label{s.prior}

\begin{figure}[!ht]
      \includegraphics[keepaspectratio,width=17cm]{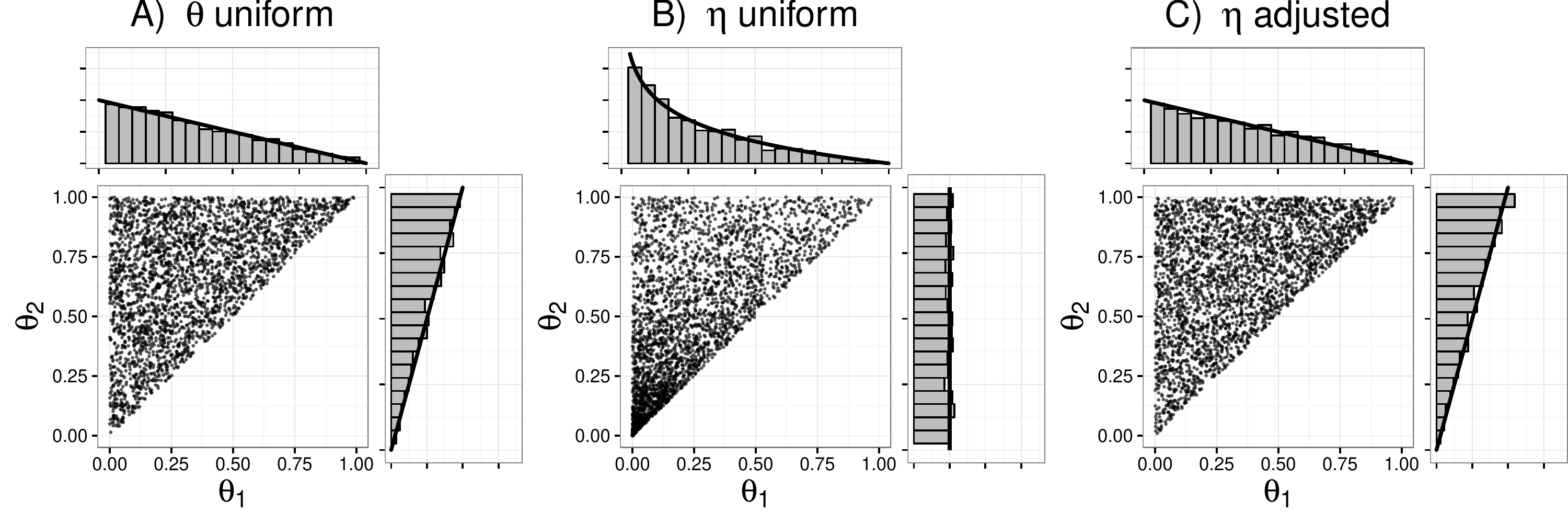}   
       \caption{Assumed (Panel A) and implied (Panel B and C) prior distributions on the parameters of interest $\bm \theta$ for a two-dimensional model \citep[$\theta_1\leq \theta_2$; cf.][Ch. 9]{lee2014bayesian-1}.} 
       \label{f.prior2d}
\end{figure}

The balanced model $\mathcal M_u^\theta$ and the unbalanced model $\mathcal M_u^\eta$ imply different prior distributions on the parameters of interest $\bm \theta$ \citep[cf. ][Ch.  9]{lee2014bayesian-1}. To illustrate the difference, we sampled parameters from the prior distributions of both models and plotted the resulting distributions of $\bm \theta$. For the reparameterized model, this requires the transformation $\theta_i = \prod_{j=i}^{P}\eta_j$ (e.g., $\theta_1=\eta_1 \cdot \eta_2\cdot\eta_3$, $\theta_2=\eta_2\cdot\eta_3$, and $\theta_3=\eta_3$). For $P=2$, Panels A and B of Figure~\ref{f.prior2d} show the resulting samples of $\bm \theta$ for each model. Whereas the balanced model puts equal probability mass on all admissible parameter values $\theta_1 \leq \theta_2$, the unbalanced model puts more mass on small values of $\theta_1$. Moreover, the balanced model has symmetric marginal distributions whereas the unbalanced model has a uniform distribution on the larger parameter $\theta_2$ and overweights small values of $\theta_1$.

This discrepancy becomes more severe as the number $P$ of order constraints increases. Since the differences in higher dimensions are difficult to visualize, we instead compare the univariate marginal prior distributions on $\theta_i$. For the balanced model $\mathcal M_u^\theta$, \citet{goggans2007assigning} showed that the marginal prior on $\theta_i$ is given by
\begin{equation}
\pi_u^{\theta_i}(\theta_i) \propto \theta_i^{i-1} (1-\theta_i)^{P-i},
\end{equation}
which is a Beta$(i, P-i+1)$ distribution. In contrast, the implied marginal prior on $\theta_i$ in the unbalanced model $\mathcal M_u^\eta$ (see Appendix \ref{a.prior}) is
\begin{equation}
\pi^{\eta_i}(\theta_i ) = \frac{1}{(P-i)!} (-\log \theta_i)^{P-i}.
\end{equation} 

\begin{figure}[!ht]
      \includegraphics[keepaspectratio,width=13cm]{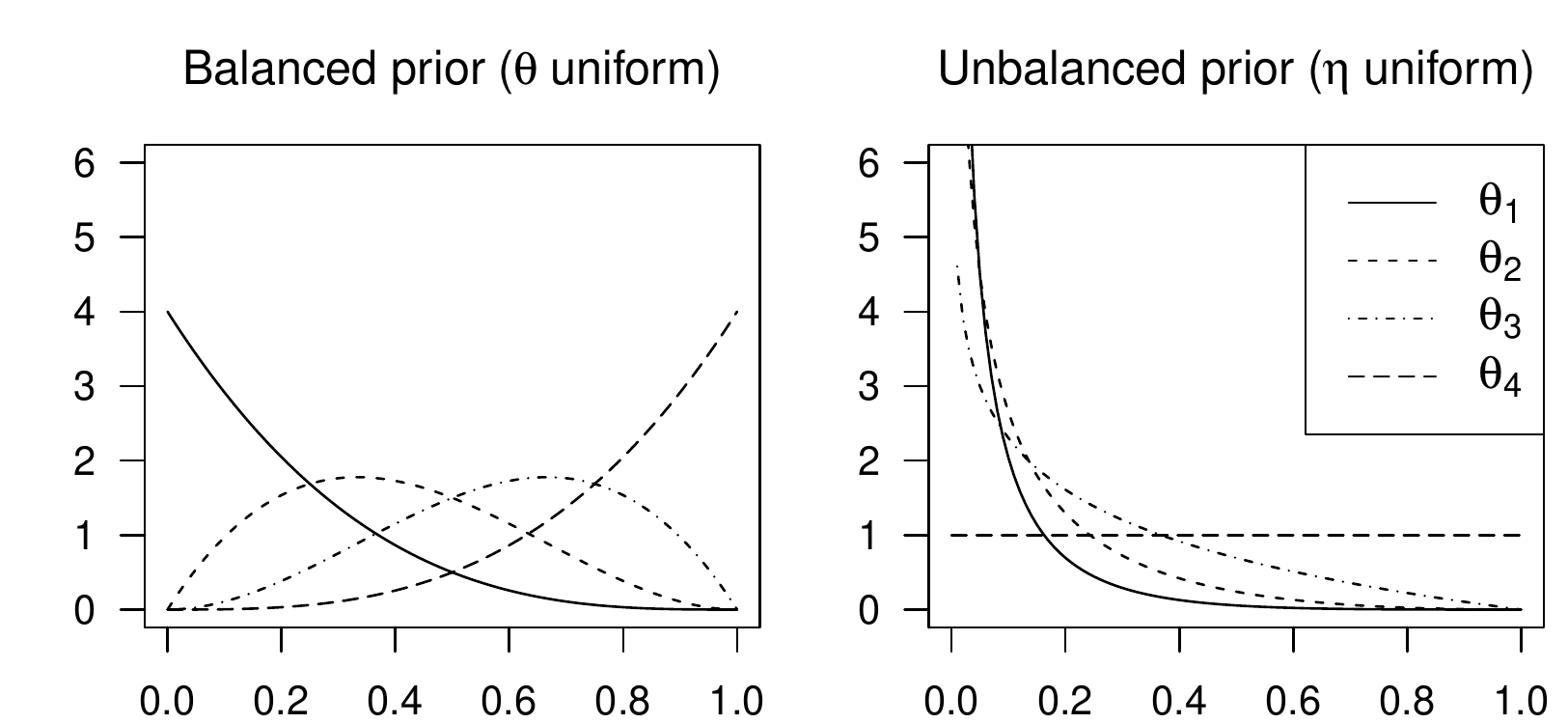}
       \caption{Marginal prior distributions on the parameters of interest $\bm \theta$ for a four-dimensional model ($\theta_1\leq \theta_2\leq \theta_3\leq \theta_4$).} 
       \label{f.prior1d}
\end{figure}

A comparison of $P=2$ in Figure~\ref{f.prior2d} and $P=4$ in Figure~\ref{f.prior1d} shows that the difference in marginal priors on $\bm\theta$ increases with the number of order constraints. The balanced prior model is symmetric in a sense that small values of the smallest parameter $\theta_1$ get as much prior mass as the complementary values of the largest parameter $\theta_P$. In contrast, the unbalanced prior model always assumes a uniform distribution on the largest parameter $\theta_P$ and overweights small values of all other parameters. Note that this asymmetry implies that any inferences based on the unbalanced model will dependent on the definition of the success probabilities $\bm\theta$ (e.g., whether to count heads or tails when tossing a coin). Most importantly, these different priors on $\bm\theta$ will have a substantial impact on the marginal likelihood, since they result in a different weighting of the parameter space when averaging the likelihood (a detailed comparison can be found in the supplementary material).

\section{Adjusted Priors for the Auxiliary Parameters} 

To ensure that a reparameterized model is equivalent to the original one, the priors for the auxiliary parameters need to be adjusted. For this purpose, we rewrite the reparameterization of $\theta_1 \leq \dots \leq \theta_P$ introduced above (Method A of Knapp \& Batchelder, \citeyear{knapp2004representing}) by the one-to-one transformation function $\bm\eta = \bm g(\bm\theta)$,
\begin{align*}
\label{e.repar}
\bm g: \Theta' &\rightarrow [0,1]^P  \numberthis \\ 
    g_i(\bm \theta) &=\theta_{i} / \theta_{i+1} \text{ for } i=1,\dots,P-1\\ 
	g_P(\bm\theta) &= \theta_P,
\end{align*}
where $\Theta'$ is the constrained parameter space, $\Theta'=\{\bm\theta \in [0,1]^P : \theta_1 \leq \dots \leq \theta_P\}$. To recover the original parameters $\bm\theta$, the inverse of this function is required:
\begin{align*}
\bm g^{-1}: [0,1]^P &\rightarrow \Theta'  \numberthis \\ 
g^{-1}_i(\bm \eta) &=\prod_{j=i}^P \eta_j \text{ for } i=1,\dots,P.
\end{align*}

In order to obtain adjusted priors for the auxiliary parameters $\bm \eta=\bm g(\bm \theta)$, we need to  apply the changes-of-variables theorem to the marginal likelihood integral in Eq.~\ref{e.marginal},
\begin{align*}
p(\bm y \mid \mathcal M_u^\theta)
&=\int_{\Theta'} f^\theta(\bm y \mid \bm \theta) \pi_u^\theta(\bm\theta) \,\mathrm d\bm \theta \\
&=\int_{\bm g(\Theta')}  f^\theta(\bm y \mid \bm g^{-1} (\bm\eta))\: P! \,|\det D\bm g^{-1}(\bm \eta)| \,\mathrm d\bm \eta \\
&=\int_{[0,1]^P}  f^\eta(\bm y \mid \bm  \eta)\, P! \,|\det D\bm g^{-1}(\bm \eta)| \,\mathrm d\bm \eta  \numberthis,
\label{e.transform}
\end{align*}
where $D\bm g^{-1}(\bm \eta)$ is the Jacobian matrix of partial derivatives of the inverse transformation function $\bm g^{-1}$. Note that this requires that the transformation function $\bm g$ is one-to-one and continuously differentiable. For the reparameterization $\bm g$ in Eq.~\ref{e.repar},
\begin{equation}
\big[D\bm g^{-1}(\bm \eta)\big]_{i,k} = \frac{\partial g^{-1}_i(\bm \eta)}{\partial \eta_k} = 
\begin{dcases}
     0, \text{ if } k<i\\
     1, \text{ if } i=k=P\\
     \prod_{\substack{j=i,\dots,P\\j\neq k}}\eta_j \, , \text{ else}
   \end{dcases}
\end{equation}
For $P=3$ parameters, this matrix is as follows:
\begin{equation*}
\begin{pmatrix}
\eta_2\eta_3 & \eta_1 \eta_3 & \eta_1 \eta_2 \\
0 & \eta_3 & \eta_2 \\
0 & 0 & 1
\end{pmatrix}.
\end{equation*}

\begin{figure}[!ht]
      \includegraphics[keepaspectratio,width=12cm]{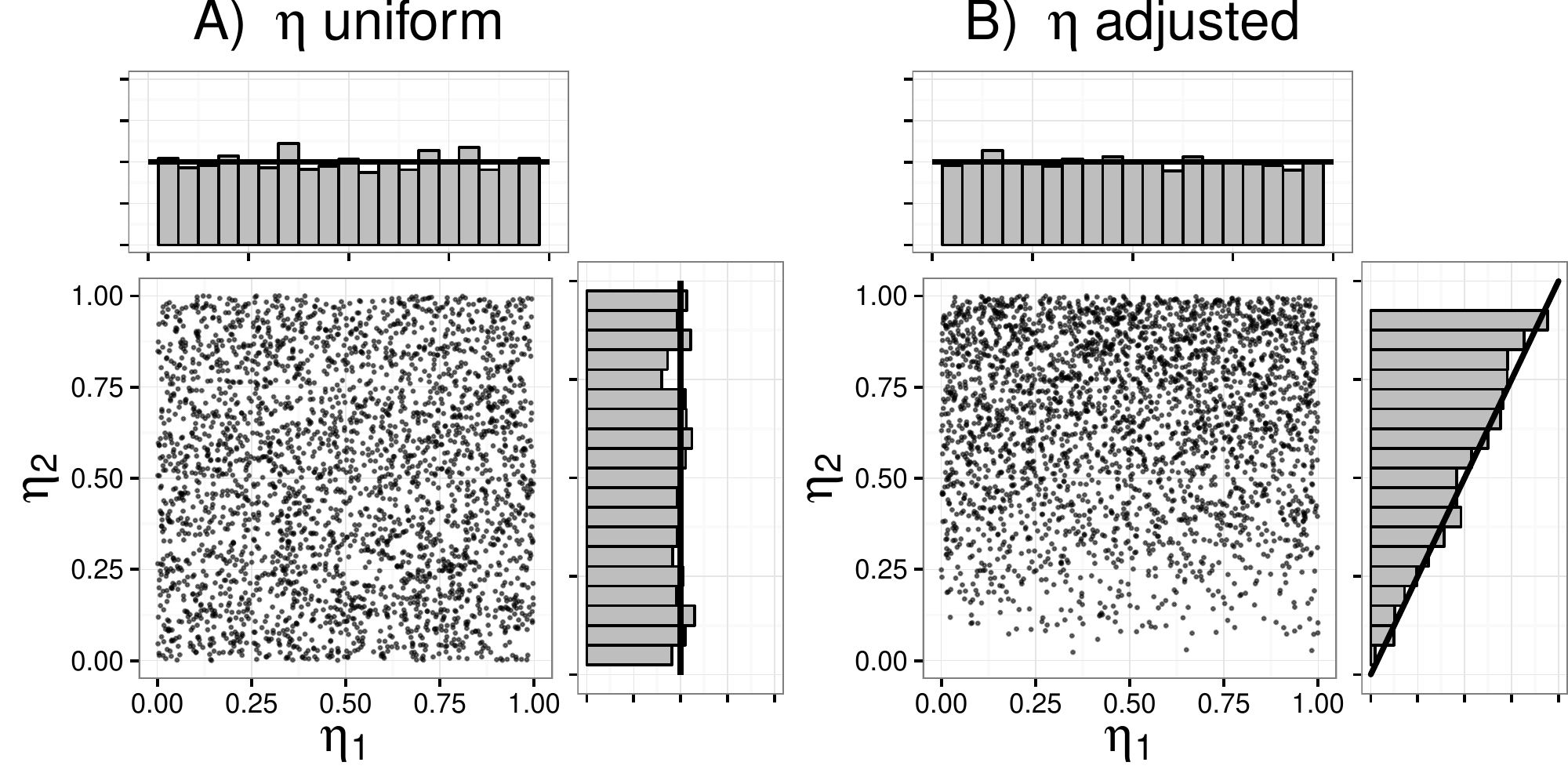}   
       \caption{Prior distributions on the auxiliary parameters $\bm \eta$ of the reparameterized, order-constrained model $\theta_1\leq \theta_2$ (cf. Figure~\ref{f.prior2d}).} 
       \label{f.prior2deta}
\end{figure}

Because all off-diagonal elements for $k<i$ are zero, the determinant of $D\bm g^{-1}(\bm \eta)$ is simply given by the product of the diagonal elements, which are always positive. Hence, based on Eq.~\ref{e.transform}, it follows that the adjusted prior for the auxiliary parameters $\bm \eta$ is
\begin{align*}
\pi_a^\eta(\bm \eta ) 
&=P! |\det D\bm g^{-1}(\bm \eta)|\\
&=P! \prod_{i=1}^{P-1} \left( \prod_{j=i+1}^{P}\eta_j\right)\\
&=P!  \prod_{i=2}^{P} \eta_i^{i-1} \\
&= \prod_{i=1}^{P}\frac{1}{\text{B}(i,1)} \eta_i^{i-1} (1-\eta_i)^{0},  \numberthis
\label{e.adjprior}
\end{align*}
which is a product of beta-distribution densities. Hence, to obtain a marginal likelihood which matches that of the balanced model with a uniform prior on the restricted parameter space, the reparameterized model requires independent beta distributions $\eta_i \sim \text{Beta}(i,1)$ as adjusted priors for the auxiliary parameters. For the two-dimensional case $P=2$, the different priors on $\bm\eta$ are illustrated in Figure~\ref{f.prior2deta}. Obviously, the adjusted prior puts more mass on large values of $\eta_2$ compared to the uniform prior, which in return implies a uniform prior on the original parameter space (Panel C of Figure~\ref{f.prior2d}).

Note that this result can easily be generalized to other statistical models with bounded parameter spaces $a\leq \theta_i \leq b$ because only the support of the integral and the corresponding scaling constants in Eq.~\ref{e.transform} change. More generally, this derivation might serve as an example how to derive adjusted priors for any type of continuously differentiable, one-to-one reparameterization. The key insight is Eq.~\ref{e.transform}: when transforming the original parameters in the marginal likelihood integral, the changes-of-variables theorem allows to obtain an adjusted prior for the new parameters.


\section{An Empirical Example: The Pair-Clustering Model}

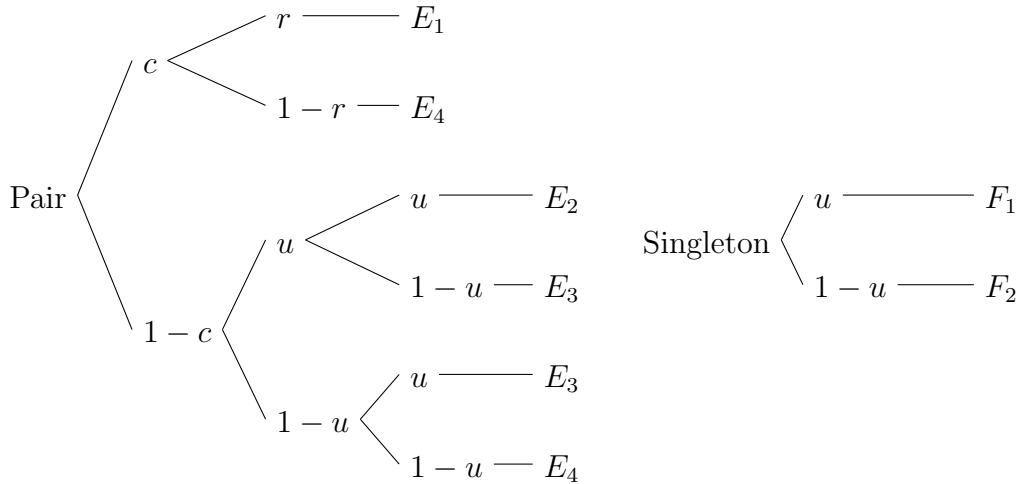
\begin{figure}[ht]
\begin{minipage}[]{0.5\linewidth}
\begin{tikzpicture}[grow=right]
    \tikzset{grow'=right,level distance=50pt}
    \tikzset{execute at begin node=\strut}
    \tikzset{every tree node/.style={anchor=base west}}
    \Tree [.Pair [.$c$   [.$r$ $E_1$ ]  [.$1-r$ $E_4$ ] ]
		         [.$1-c$ 
		             [.$u$ [.$u$ $E_2$ ] [.$1-u$ $E_3$ ]]
				     [.$1-u$ [.$u$ $E_3$ ] [.$1-u$ $E_4$ ]]]]
\end{tikzpicture}
\end{minipage}
\hspace{0.1cm}
\begin{minipage}[]{0.2\linewidth}
\begin{tikzpicture}[grow=right]
    \tikzset{grow'=right,level distance=64pt}
    \tikzset{execute at begin node=\strut}
    \tikzset{every tree node/.style={anchor=base west}}
    \Tree [.Singleton [.$u$   $F_1$ ] [.$1-u$ $F_2$ ]]
\end{tikzpicture}
\end{minipage}
\caption{The pair-clustering model \citep{batchelder1986statistical}. See text for details.}
\label{f.pc}
\end{figure}

The following example of pair clustering in memory \citep{batchelder1986statistical, batchelder1980separation} highlights the importance of adjusted priors. The standard paradigm consists of a learning and a free-recall phase and includes words that are semantically associated (pairs) and words that are semantically unrelated (singletons). To account for the finding that pairs are usually recalled either consecutively or not at all, \citet{batchelder1980separation} proposed the pair-clustering model shown in Figure~\ref{f.pc}. Given a pair of semantically associated words, both words are stored jointly with probability $c$. If storage was successful, the whole cluster is retrieved from memory with probability $r$ resulting in consecutive recall of both words (category $E_1$). In contrast, if retrieval fails with probability $1-r$, neither word is recalled ($E_4$). In case of unsuccessful clustering of a pair, both words can only be stored and recalled independently with probability $u$ resulting either in non-consecutive recall of both items ($E_2$), in recall of one of the two items ($E_3$), or in recall of neither item ($E_4$). Similarly, singletons can only be stored and retrieved separately with probability $u$ resulting either in recall ($F_1$) or not ($F_2$). 

\begin{figure}[!ht]
      \includegraphics[keepaspectratio,width=15cm]{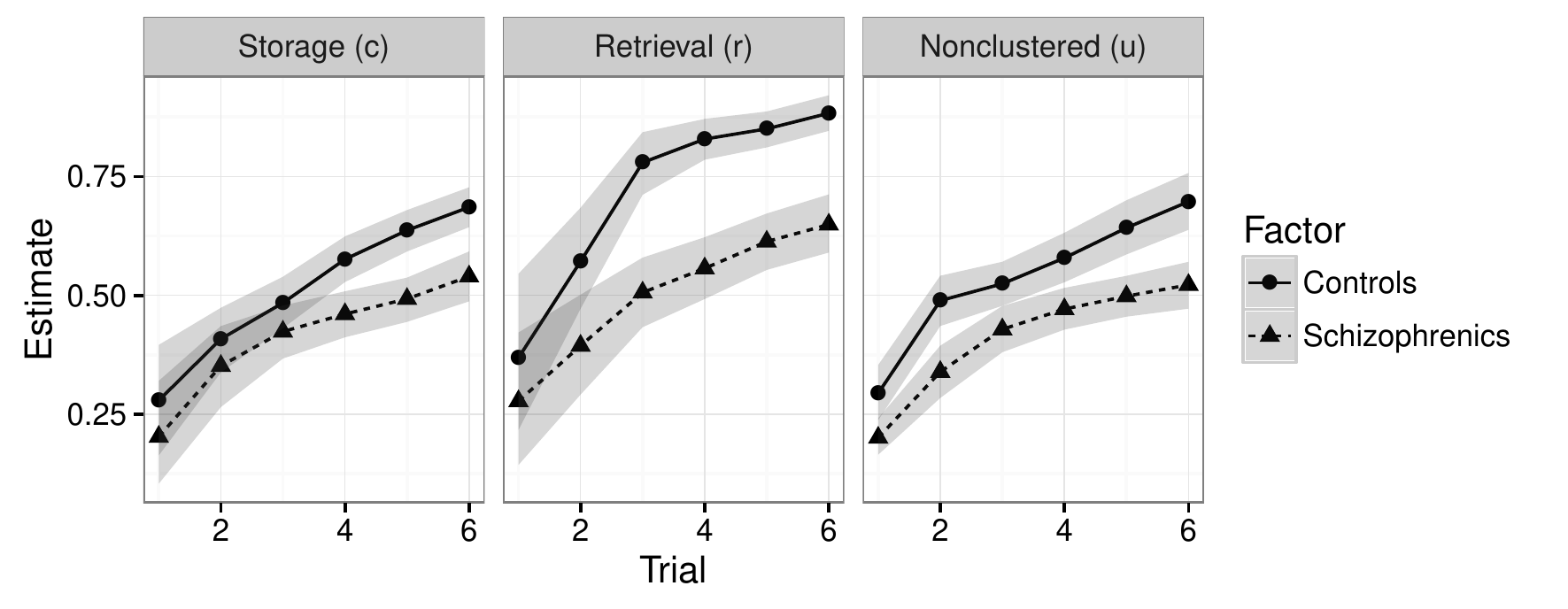}
       \caption{Mean posterior estimates and 95\% credibility intervals for Experiment 3 of \citet{riefer2002cognitive} based on the order-constrained model $\mathcal M_a^\eta$ with adjusted priors on the auxiliary parameters.} 
       \label{f.riefer}
\end{figure}

\citet{riefer2002cognitive} used the pair-clustering model to test the specificity of memory deficits in schizophrenic patients. For this purpose, six study-test trials were administered to compare the memory performance of healthy controls and schizophrenics. \citet{riefer2002cognitive} predicted that the three memory parameters for storage ($c$), retrieval ($r$), and single-item memory ($u$) should increase over trials due to learning. For the storage parameter $c$, this hypothesis is represented by the order constraints $ c_{i1}  \leq  \dots \leq c_{i6}$ within each group $i=1,2$ (the same order constraints hold for $r_{ik}$ and $u_{ik}$). In line with this prediction, the posterior estimates, shown in Figure~\ref{f.riefer}, indeed indicate an increase of memory performance across trials in both groups. 

We used the importance sampling approach by \citet[][]{vandekerckhove2015model}\footnote{See the supplementary material for a detailed explanation and implementation in R.} to compute the marginal likelihoods of four models: the balanced, order-constrained model $\mathcal M_a^\eta$ with a uniform prior on the original parameter space (reparameterized with adjusted priors); the unbalanced, reparameterized model $\mathcal M_u^\eta$ with uniform auxiliary parameters; the full model $\mathcal M_f^\theta$ without constraints on $c_{ij}$, $r_{ij}$, and $u_{ij}$; and the null model $\mathcal M_n^\theta$ with only three parameters $c_{i}$, $r_{i}$, and $u_{i}$ that are constant across study-test trials for each group $i=1,2$. 

Table~\ref{t.bf} shows that the balanced model $\mathcal M_a^\eta$ performed best by several orders of magnitude. Importantly, the difference in marginal likelihoods between the models $\mathcal M_a^\eta$ and $\mathcal M_u^\eta$ was considerable as indicated by a log Bayes factor of $21.06$, or equivalently, a Bayes factor of around 1.4 billion. This huge discrepancy is noteworthy because both models implement the same order constraints and differ only in the prior distribution on the restricted parameter space. Obviously, there is much more evidence for a uniform distribution on the order-constrained, original parameter space as represented by the model $\mathcal M_a^\eta$. The Bayes factor for the two order-constrained models is even larger than that comparing the reparameterized model $\mathcal M_a^\eta$ against the full model $\mathcal M_f^\theta$. Hence, in terms of evidence, the choice of priors for the auxiliary parameters weighs stronger than the order constraint itself.

\begin{table}[!ht]
\centering
\caption{Estimated marginal likelihoods for Experiment 3 of \citet{riefer2002cognitive}.}
\label{t.bf}
\begin{tabular}{lrrrr}
  \hline
 \multicolumn{1}{c}{Model $\mathcal M$} & 
 \multicolumn{1}{c}{$p(\bm y \mid \mathcal M)$ }& 
 \multicolumn{1}{c}{SE$_p$ }& 
 \multicolumn{1}{c}{$\log p(\bm y \mid \mathcal M)$} & 
 \multicolumn{1}{c}{SE$_{\log p}$ }\\ 
  \hline  
Full & 7.26\e{-87} & 1.45\e{-87} & $-198.68$ & $0.10$ \\ 
Balanced & 3.30\e{-74} & 2.55\e{-75} & $-169.30$ & $0.06$ \\
Unbalanced & 2.64\e{-83} & 3.58\e{-84} & $-190.36$ & $0.08$ \\ 
Null & 6.98\e{-264} & 1.06\e{-266} & $-605.94$ & $ <0.01$ \\ \hline
\end{tabular}
\end{table}


\section{Discussion}

Order constraints are important for testing psychological theories and are often implemented by reparameterizing the original parameters, especially in MPT modeling \citep{knapp2004representing}. However, whereas inferences based on maximum likelihood are not affected by reparameterization, care has to be taken when relying on Bayesian model selection. Using a simple product-binomial model, we illustrated the difference between a uniform prior on the order-constrained, original parameters versus a uniform prior on the auxiliary parameters of the reparameterized model. Whereas the former puts equal prior mass to all of the admissible original parameters, the latter puts more weight on small parameter values (Figure~\ref{f.prior2d}).  We derived adjusted priors for the auxiliary parameters (beta distributions) that result in identical marginal likelihoods as those obtained from the original model. In practice, the Bayes factor can differ substantially if the prior does not take the reparameterization into account as illustrated in the pair-clustering example \citep{riefer2002cognitive}. 

The impact of priors shown in the present paper might support previous criticism of the Bayes factor due to its prior sensitivity \citep[e.g.,][]{liu2008bayes}. However, the prior sensitivity can be seen as one of the major advantages of Bayesian model selection \citep[e.g.,][]{vanpaemel2010prior}. By incorporating prior knowledge into a cognitive model, the model becomes more specific and makes stronger predictions. In line with this view, order constraints can be interpreted as highly informative priors that put zero probability on large proportions of the original parameter space \citep{lee2016bayesian}. Importantly, constraining the prior probability to be zero is qualitatively different from assuming very low but nonzero probability mass. If the prior is zero on some parameter subspace, the posterior will always be zero on this subspace regardless of the data. In contrast, for an arbitrarily small but nonzero prior probability, it is in principle possible to observe data that substantially increase the posterior probability on this subspace. Therefore, if the Bayes factor is criticized for its sensitivity with respect to the prior distribution, the same critique applies to order constraints in general, which merely encode highly informative prior knowledge about the parameters.

\subsection{Conclusion}

The Bayes factor has many advantages that are desirable for model selection \citep{wagenmakers2007practical} --- it directly emerges from the application of Bayes' theorem when computing posterior model odds, it has a direct interpretation as the evidence in favor of one model versus another, and it considers order constraints and the functional complexity of statistical models \citep{myung1997applying}. However, if order constraints are reparameterized, priors on the new auxiliary parameters need to be adjusted to ensure meaningful priors on the original parameters of interest. For linear order constraints in MPT models, we showed that these adjusted priors are independent beta distributions. With the present paper, we hope to raise awareness about this issue and show the severe consequences of using default priors for reparameterized models.

\printbibliography
\appendix


\section{Marginal Priors for the Reparameterized Model}
\label{a.prior}

We derive the marginal priors $\pi^{\theta_i}$ on the original parameters $\theta_i$ implied by the unbalanced prior model with uniform prior distributions for the auxiliary parameters $\bm\eta$. More precisely, given the prior distribution
\begin{equation}
\pi^\eta_u(\bm \eta ) = \prod_{i=1}^{P} \mathcal I_{[0,1]}(\eta_i),
\end{equation}
it follows that the original parameters 
$\theta_i = \prod_{j=i}^P \eta$ have the marginal distribution
\begin{equation}
\pi^{\theta_i}(\theta_i) = \frac{1}{(P-i)!} (-\log \theta_i)^{P-i}
\label{e.IA}
\end{equation}
for $\theta_i \in [0,1]$. We prove this by backward induction. Obviously, $\theta_P=\eta_P \in [0,1]$ and hence 
\begin{align*}
\pi^{\theta_P}(\theta_P ) = \mathcal I_{[0,1]}(\theta_P) =  \frac{1}{(P-P)!} (-\log \theta_P)^{P-P}.
\end{align*}
Next, assume that Eq.~\ref{e.IA} holds for $i \in\{2,\dots,P\}$. Then, for $i \rightarrow  i-1$, the cumulative density function of $\theta_{i-1}$ is given by
\begin{align*}
F^{\theta_{i-1}}(x) 
&=\Pr\left[\eta_{i-1}\theta_i  \leq x\right]\\
&= \int_{0}^1 \Pr[\eta_{i-1} \leq \frac x t] \,\, \pi_r^{\theta_i}(t) \,\mathrm d t\\
&= \frac{1}{(P-i)!} \left[\int_{0}^x 1 \cdot  (-\log t)^{P-i} \mathrm d t + x\int_{x}^1  \frac 1 t (-\log t)^{P-i} \mathrm d t\right]
\end{align*}
Taking the derivative $\mathrm d/\mathrm dx$ of the cumulative density function results in
\begin{align*}
\pi^{\theta_{i-1}}(x) &=
\frac{1}{(P-i)!}\left[(-\log x)^{P-i} + \int_{x}^1  \frac 1 t (-\log t)^{P-i} \mathrm d t - x \cdot\frac 1 x (-\log x)^{P-i}\right]\\
&= \frac{1}{(P-i)!}  \int_{x}^1  \frac 1 t (-\log t)^{P-i} \mathrm d t\\
&=\frac{1}{(P-(i-1))!} (-\log x)^{P- (i-1)},
\end{align*}
which completes the proof.

\section{Adjusted Priors for Method B of Knapp \& Batchelder (2004)}
\label{a.methodb}

\subsubsection{Method B}

Whereas the auxiliary parameters $\bm \eta$ in the main text are defined as the proportional decreases of the original parameters $\theta_P$, $\theta_{P-1}$,... $\theta_1$, order constraints can also be defined in terms of proportional increases of $\theta_1$, $\theta_2$,... $\theta_P$. In this approach, called Method B by \citet{knapp2004representing}, the continuously differentiable, one-to-one transformation function $\bm \eta=\bm g(\bm\theta)$ is given by 
\begin{align*}
	\begin{dcases}
g_1(\bm\theta) = \theta_1\\
g_i(\bm\theta) = \frac{\theta_i- \theta_{i-1}}{1-\theta_{i-1}},
   \end{dcases}
\end{align*}
in which the parameters $\eta_i$ can be interpreted as `growth parameters.' The inverse is given by
\begin{equation}
g^{-1}_i(\bm \eta) =  1-\prod_{j=1}^i(1-\eta_j) \, ,
\end{equation}
Moreover, the Jacobian $D\bm g^{-1}(\bm \eta)$ is
\begin{equation}
\left[D\bm g^{-1}(\bm \eta)\right]_{i,k} =\frac{\partial g^{-1}_i(\bm \eta)}{\partial \eta_k} = 
\begin{dcases}
     0, \text{ if } k<i\\
     1, \text{ if } i=k=1\\
     \prod_{\substack{j=1,\dots,i\\j\neq k}}(1-\eta_j) \, , \text{ else}.
   \end{dcases}
\end{equation}
The determinant of this triangular matrix is the product of the diagonal elements, which leads to the adjusted prior
\begin{align*}
\pi_a^{\eta}(\bm \eta) 
&= P! \,|\det D\bm g^{-1}(\bm \eta)|\\
&= P! \,\prod_{i=2}^{P} \left( \prod_{j=1}^{i-1}(1-\eta_j)\right)\\
&= P!  \prod_{i=1}^{P-1} (1-\eta_i)^{P-i} \\
&= \prod_{i=1}^{P} \frac{1}{\text{B}(1,P-i+1)}\eta_i^{0} (1-\eta_i)^{P-i}. \numberthis
\end{align*}
It follows that the adjusted prior for the reparameterized model requires independent beta priors on the auxiliary parameters, that is, $\eta_i \sim \text{Beta}(1,P-i+1)$.

\end{document}